\documentclass[]{aa}
\bibpunct{(}{)}{;}{a}{}{,}
\usepackage{color}
\usepackage[varg]{txfonts}

\newcommand{\teff}{$T_\mathrm{eff}$}
\newcommand{\vt}{$V_\mathrm{tan}$}
\usepackage{graphicx}

\usepackage{txfonts}

\begin{document}

   \title{J-PLUS: Discovery and characterisation of ultracool dwarfs \\ using Virtual Observatory tools}

   %\subtitle{I.}

   \author{E.~Solano\inst{1,2}
\and
E.~L.~Mart\' {\i}n \inst{3}
\and 
J.~A.~Caballero\inst{1}
\and
C.~Rodrigo\inst{1,2}
\and
R.~E.~Angulo\inst{4,5,6}
\and
J.~Alcaniz\inst{7}
\and
M.~Borges Fernandes\inst{7}
\and
A.~J.~Cenarro\inst{8}
\and
D.~Crist\'{o}bal-Hornillos\inst{8}
\and
R.~A.~Dupke\inst{7,9,10,11}
\and
E.~Alfaro\inst{11}
\and
A.~Ederoclite\inst{8,12}
\and
F.~Jim\'{e}nez-Esteban\inst{1,2}
\and
J.~A.~Hernandez-Jimenez\inst{12,13}
\and
C.~Hern\'{a}ndez-Monteagudo\inst{8}
\and 
R.~Lopes de Oliveira\inst{14,15,7}
\and
C.~L\'opez-Sanjuan\inst{8}
\and
A.~Mar\' {\i}n-Franch\inst{8}
\and
C.~Mendes~de~Oliveira\inst{12}
\and
M.~Moles\inst{4}
\and
A.~Orsi\inst{4}
\and
L.~Schmidtobreick\inst{16}
\and
D.~Sobral\inst{17}
\and
L.~Sodr\'{e} Jr.\inst{12}
\and
J.~Varela\inst{8}
\and
 H.~V\'{a}zquez Rami\'{o}\inst{4}
}
   
  \institute{
  Departamento de Astrof\'{\i}sica, Centro de Astrobiolog\'{\i}a (CSIC-INTA), ESAC Campus, Camino Bajo del Castillo s/n, E-28692, Villanueva de la Ca\~{n}ada, Madrid, Spain
\and 
Spanish Virtual Observatory, Spain
\and 
Departmento de Astrof\'{\i}sica, Centro de Astrobiolog\'{\i}a (CSIC-INTA), Carretera de Torrej\'{o}n a Ajalvir km 4, E-28850 Torrej\'{o}n de Ardoz, Madrid, Spain
\and
Centro de Estudios de F\'{\i}sica del Cosmos de Arag\'{o}n (CEFCA), Plaza San Juan 1, 44001 Teruel, Spain
\and
Donostia International Physics Centre (DIPC), Paseo Manuel de Lardizabal 4, E-20018 Donostia-San Sebasti\'{a}n, Spain
\and
IKERBASQUE, Basque Foundation for Science, E-48013, Bilbao, Spain
\and
Observatorio  Nacional - MCTIC (ON), Rua General Jos\'{e} Cristino, 77, S\~{a}o Crist\'{o}v\~{a}o, 20921-400, Rio de Janeiro,  Brazil
\and
Centro de Estudios de F\'{\i}sica del Cosmos de Arag\'{o}n (CEFCA) -- Unidad Asociada al CSIC, Plaza San Juan 1, 44001 Teruel, Spain
\and 
University of Michigan, Department of Astronomy, 1085 South University Ave., Ann Arbor, MI 48109, USA
\and
University of Alabama, Department of Physics and Astronomy, Gallalee Hall, Tuscaloosa, AL 35401, USA
\and
Instituto de Astrof\'{\i}sica de Andaluc\'{\i}a, Glorieta de la Astronom\'{\i}a s/n, 18008 Granada, Spain
\and
Instituto de Astronomia, Geof\'{\i}sica e Ciências Atmosféricas, Universidade de S\~{a}o Paulo, Rua do Mat\~{a}o 1226, C. Universitária, S\~{a}o Paulo, 05508-090, Brazil
\and
Departamento de Ciencias F\'{\i}sicas, Universidad Andr\'{e}s Bello, Fern\'{a}ndez Concha 700, Las Condes, Santiago, Chile
\and
Departamento de F\'{\i}sica, Universidade Federal de Sergipe, Av. Marechal Rondon, s/n.  S\~{a}o Crist\'{o}v\~{a}o, 49100-000 Sergipe (SE), Brazil
\and
Center for Space Science and Technology, University of Maryland, Baltimore County, 1000 Hilltop Circle, Baltimore, MD 21250, USA
\and
European Southern Observatory. Alonso de Cordova 3107, Vitacura, Santiago, Chile.
\and
 Department of Physics, Lancaster University, Lancaster LA1 4YB, United Kingdom
}

   \date{Received 12 02 2019 / Accepted  07 05 2019}

% \abstract{}{}{}{}{} 
% 5 {} token are mandatory
 
  \abstract
   {Ultracool dwarfs are objects with spectral types equal or later than M7. Most of them have been discovered using wide-field imaging surveys. The Virtual Observatory has proven to be of great utility to efficiently exploit these astronomical resources}
   {We aim to validate a Virtual Observatory methodology designed to discover and characterize ultracool dwarfs in the J-PLUS photometric survey. J-PLUS is a multiband survey carried out with the wide angle T80Cam optical camera mounted on the 0.83-m telescope JAST/T80 in the Observatorio Astrof\' {\i}sico de Javalambre. In this work we make use of the Internal Data Release covering 528\,deg$^2$.}
   {We complement J-PLUS photometry with other catalogues in the optical and infrared using VOSA, a Virtual Observatory tool that estimates physical parameters from the spectral energy distribution fitting to collections of theoretical models. Objects identified as ultracool dwarfs are distinguished from background M giants and highly reddened stars using parallaxes and proper motions from {\it Gaia} DR2. }
   {We identify 559 ultracool dwarfs, ranging from $i$=16.2 mag to $i$=22.4 mag, of which 187 are candidate ultracool dwarfs not previously reported in the literature. This represents an increase in the number of known ultracool dwarfs of about 50\% in the studied region of the sky, particularly at the faint end of our sensitivity, which is interesting as reference for future wide and deep surveys such as Euclid. Three candidates constitute interesting targets for exoplanet surveys because of their proximity (distances less than 40 pc).  We also analyze the kinematics of ultracool dwarfs in
our catalogue and find evidence that it is consistent with a Galactic thin-disk population, except for six objects that might be members of the thick disk.}
   {The results obtained in this paper validate the proposed methodology, which will be used in future J-PLUS and J-PAS releases. Considering the region of the sky covered by the Internal Data Release used in this work, we foresee to discover 3\,000-3\,500 new ultracool dwarfs at the end of the J-PLUS project.}

   \keywords{astronomical data bases: surveys -- astronomical data bases: virtual observatory tools -- stars: low-mass -- brown dwarfs.
               }

   \maketitle
%
%________________________________________________________________
\section{Introduction}
Ultracool dwarfs (UCDs)  are defined as objects with spectral types equal or later than M7\,V. They comprise both the lowest mass hydrogen-fusing stars and brown dwarfs (substellar objects not massive enough to sustain steady hydrogen fusion). The M7\,V spectral type marks the beginning of a variety of changes with decreasing effective temperature, in particular the appearance of dust clouds, which makes an increasingly important contribution to the atmospheric chemistry \citep[e.g.,][]{Jones97, Helling08}. UCDs represent about 15\% of the stellar population in the solar neighborhood \citep[e.g.,][]{Henry2006, Gillon16}

While the ages of UCDs favor studies of Galactic kinematics, they also play a relevant role in two other types of research lines. On one hand, their low masses and small radii facilitate the discovery of Earth-size planets orbiting around them at separations at which surface water could be liquid. Moreover, their proximity to the Sun (UCDs are intrinsically faint objects that cannot be detected at long distances) opens the door to the detection and characterization of habitable planets in the solar neighbourhood. Proxima Centauri \citep{Anglada16}, TRAPPIST\,-1 \citep{Gillon17}, and Barnard's star \citep{Ribas18} are excellent examples of this type of research. Given that there are approximately three times as many M dwarfs as FGK dwarfs in the Milky Way \citep{Kroupa01, Chabrier03}, and small planets appear to surround M dwarfs three to five times more frequently than Sun-like stars \citep{Dressing15}, they could well represent the most common Earth-size planets in our Galaxy.

On the other hand, some of the youngest ($t$ $\leq$ $100$\,Myr) UCDs in the field may have masses close to or even below the deuterium burning limit (10 - 13 M$_{\rm Jup}$; \citealt {Chabrier00, Zhang19}), which makes them exoplanet analogues. Since these objects are nearby, isolated, and not affected by the glare of the host star, they are ideal laboratories for detailed studies of the cool, low-gravity and dusty atmospheres typical of exoplanets \citep{Faherty13, Caballero18}.

The major sources of UCD discoveries, which now include over 2\,000 L and T dwarfs and several thousands of late-M dwarfs \citep{Smart17}, have been wide-field, optical and infrared imaging surveys such as the Deep Near Infrared
Survey of the Southern Sky \citep{Epchtein99}, the Sloan Digital Sky Survey \citep{York00},
the Two Micron All Sky Survey \citep{Skrutskie06}, the UKIRT Infrared Deep Sky Survey \citep{Lawrence07}, the Wide-Field Infrared Survey Explorer \citep{Wright10}, and the Visible and Infrared Survey Telescope for Astronomy \citep{Cross12}. In the near future, a major source of UCD identifications is expected to be the Euclid surveys \citep{Deacon18}.

In this context, we plan to search for UCDs in J-PLUS\footnote{\tt https://www.j-plus.es/} (Javalambre Photometric Local Universe Survey) and J-PAS\footnote{\tt http://www.j-pas.org/} (Javalambre Physics of the Accelerating Universe Astrophysical Survey) taking advantage of the Virtual Observatory\footnote{\tt http://www.ivoa.net} (VO) capabilities. VO has proven to be an excellent methodology to identify and characterize cool objects \citep[e.g.,][]{Aberasturi11, Martin13, Aberasturi14, Galvez17, Lodieu17}. In this paper we present the first search using the J-PLUS Internal Data Release catalogue which covered 528\,deg$^{2}$ ($\sim$ 1\% of the total sky area) in the blue and red optical bands.

The main advantage of J-PLUS compared to the previously cited surveys is its larger number of filters, which provide a denser sampling of the spectral energy distribution (SED) and allows a more accurate estimation of the effective temperature, a key parameter to classify an object as a UCD. The fact that all photometric information comes from the same survey also minimizes the risk of mismatching among surveys conducted at different epochs, in particular for the most interesting, high proper motion, nearby sources. 

This paper is organized as follows. In Sect. 2 we describe the J-PLUS survey. Sect. 3 is devoted to explain the methodology that we have used to identify candidate ultracool dwarfs. In Sect. 4, we present some properties of our candidates.  Finally, we summarize our work and present our conclusions in Sect. 5. 
%__________________________________________________________________

\section{J-PLUS}
\label{jplus}
J-PLUS is an ongoing multi-filter survey carried out with the
Javalambre Auxiliary Survey Telescope (JAST/T80), a 0.83\,m
telescope installed at the Observatorio Astrof\' {\i}sico de Javalambre
(OAJ) in Teruel, Spain. The survey uses the panoramic camera T80Cam
that provides a large field of view of 1.4 $\times$ 1.4 deg$^{2}$ with a pixel scale of 0.55\,arcsec/pixel. J-PLUS was primarily conceived to perform the
calibration tasks for the main J-PAS survey, observing the same regions of the sky ($\approx$ 8\,500 deg$^{2}$). J-PAS \citep{Benitez14}, is a photometric sky survey in 59 colours that will allow, for the first time, to map not only the positions of hundreds of millions of galaxies in the sky but also their individual distances, providing the first complete 3D map of the Universe.

The J-PLUS filter system is composed of four broad- ($g$, $r$, $i$, and $z$), two intermediate- ($u$ and $J0861$), and six
narrow-band ($J0378$, $J0395$, $J0410$, $J0430$, $J0515$, and
$J0660$) filters optimized to provide an adequate sampling of the target SED in the optical range. The transmission curves, as well as additional properties of the set of filters, can be found at the Filter Profile Service maintained by the Spanish Virtual Observatory\footnote{\tt https://bit.ly/2OXuwNs}. The expected limiting magnitude (3$\sigma$\, in 3\,arcsec aperture) is $\sim$ 21\,mag, with $g$ and $r$ reaching $\sim$ 22\,mag. J-PLUS magnitudes are given in the AB system.

The final survey parameters, scientific goals and the technical requirements of the filter set are described
in \cite{Cenarro19}. J-PLUS data have already been used in different research fields, such as the study of the Coma cluster \citep{Teja19}, the analysis of the H$\alpha$ emission in the nearby Universe \citep{Logrono19}, the study of the M15 globular cluster \citep{Bonatto19}, the identification of new members in a galaxy cluster \citep{Molino19}, the study of the stellar populations of several local galaxies \citep{Sanroman19}, or the study of the star/galaxy morphological classification \citep{Lopez19}.

\section{Analysis}
\label{method}
\begin{figure*}
\includegraphics[width=16.4cm]{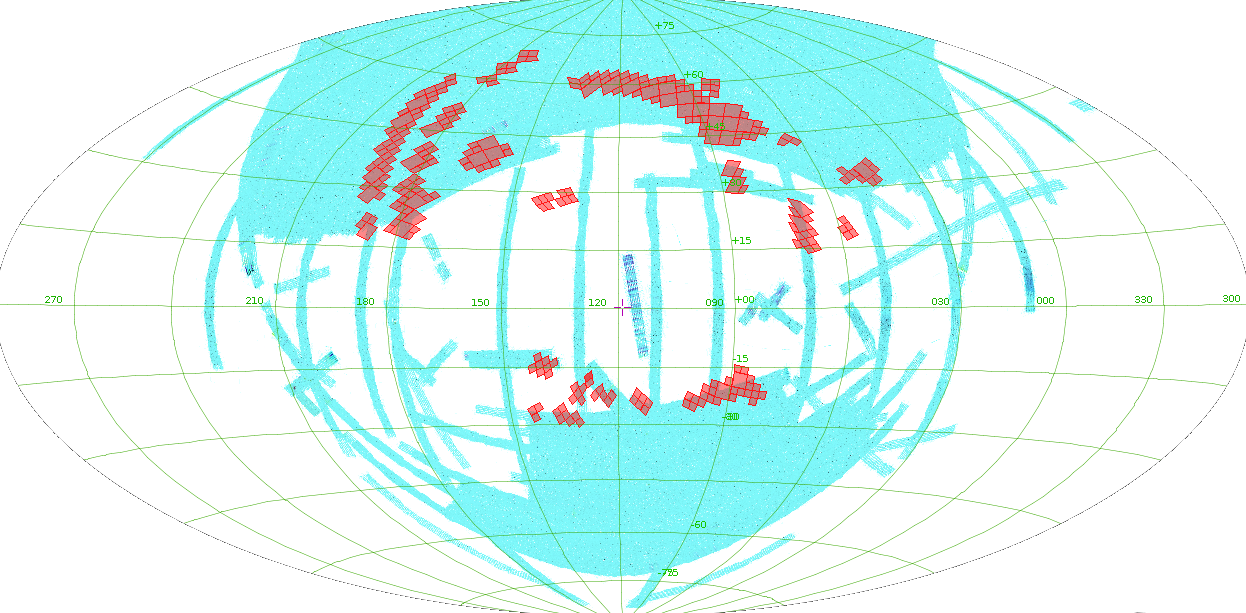}
    \caption{Sky coverage of J-PLUS IDR201709 in Galactic coordinates (red). The SDSS DR9 footprint (blue) is plotted for comparison.}
	\label{fig:aladin}
\end{figure*}

We queried the J-PLUS archive\footnote{\tt https://archive.cefca.es/catalogues}, in particular the Internal Data Release (IDR201709), using the VO Asynchronous Queries option. IDR201709 comprises 264 J-PLUS pointings observed in the 12 optical bands described in Sect.\,\ref{jplus} amounting to 528 deg$^2$ (Fig.\,\ref{fig:aladin}). These pointings, available for the J-PLUS collaboration in September 2017, are a subset of the 511 pointings that comprise the first J-PLUS data release, DR1\footnote{\tt https://www.j-plus.es/datareleases/data$_{-}$release$_{-}$dr1}, presented by \cite{Cenarro19}. Differences at 0.1\,mag level in the photometry of the same source between IDR201709 and DR1 can exist due to improvements in the reduction and the calibration processes made after September 2017. IDR201709 includes two types of catalogues: single catalogues, where the source detection and photometry were done on each band independently, and dual catalogues, where the detection and photometry were done using the $r$-band image as a reference. As UCDs emit most of their flux at longer wavelengths, we decided not to be linked to the $r$ band and use the single catalogues in our analysis, instead. For this, we crossmatched the sources extracted from the images at different bands assuming that sources separated less than 1\,arcsec are, actually, the same source. 

%As the maximum number of sources retrieved in a single query is limited to 1\,000\,000 and IDR201709 contains several million sources, we had to submit more than one query to cover the whole IDR201709 footprint.

\normalsize

\subsection{Photometric search}
\label{phot}
We restricted our query to stellar sources. This was done using the {\tt class\_star} parameter of SExtractor \citep{Bertin96}. Sources having {\tt class\_star} $>$ 0.5 in the $r$-band image are considered ``stars'' in the J-PLUS archive. In order to minimize the contamination by non-stellar sources we were more restrictive and imposed {\tt class\_star} $>$ 0.8. We also required good photometric conditions (flag=0). An example of a typical query looks like this:\\

\scriptsize
\begin{verbatim}
SELECT * FROM jplus.CalibratedMagABSingleObj,jplus.Filter 
WHERE jplus.Filter.FILTER_ID = jplus.CalibratedMagABSingleObj.FILTER_ID 
AND ALPHA_J2000 BETWEEN 145.0 AND 146.0 
AND DELTA_J2000 BETWEEN 30.0 AND 33.0 
AND CalibratedMagABSingleObj.FLAGS=0 A
AND CalibratedMagABSingleObj.CLASS_STAR>0.8
\end{verbatim}

\begin{figure}
	\includegraphics[width=\columnwidth]{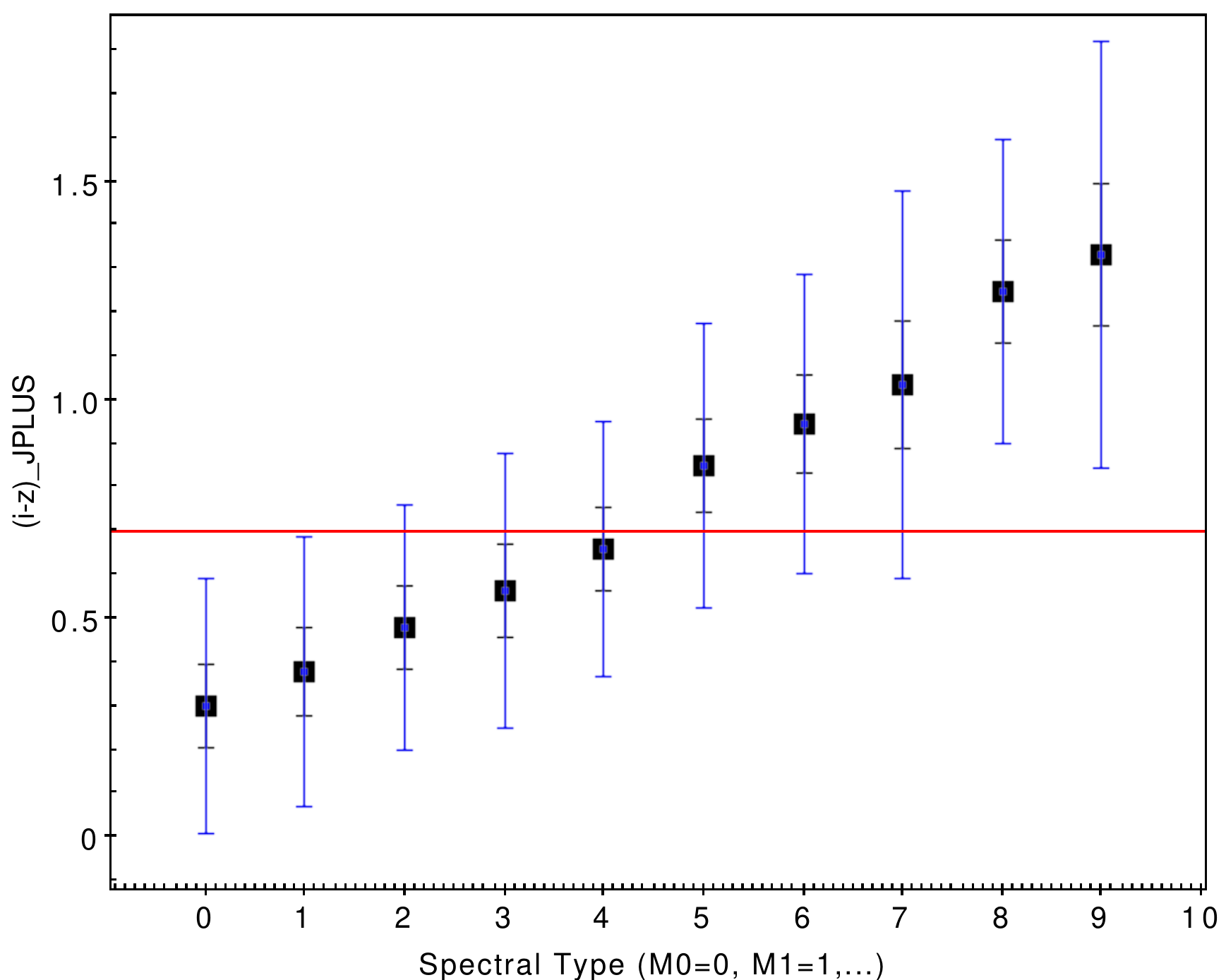}
    \caption{Mean colour $i-z$ as a function of spectral type for the M dwarfs included in \citet{West11} and with observations in J-PLUS. The red line represents the $i-z$ = 0.7\,mag colour cut. Bars at 1$\sigma$ and 3$\sigma$ are also shown.
}
	\label{fig:sptwest}
\end{figure}

\begin{figure}
	\includegraphics[width=\columnwidth]{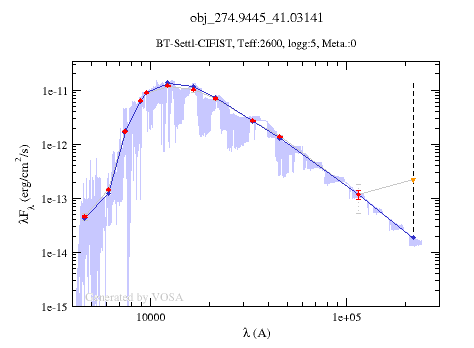}
    \caption{Example of a SED fitting as generated by VOSA. The blue spectrum represents the theoretical model that best fits while red dots represent the observed photometry. The inverted yellow triangle indicates that the photometric value corresponds to an upper limit. The vertical dashed line flags a possible excess in the SED.
}
	\label{fig:vosa}
\end{figure}

\normalsize
Figure\,\ref{fig:sptwest} shows the distribution of the $i-z$ J-PLUS colours for the list of spectroscopically confirmed M dwarfs given by \cite{West11} and included in IDR201709. A list of 1257 stars was obtained from this crossmatch (Table\,\ref{tab:teffcol}). For this subset, $i-z$ colour ranges from a mean value of $i-z$ $\sim$ 0.3\,mag at subtype M0\,V to $i-z$ $\sim$ 1.3\,mag at subtype M9\,V. We adopted a colour-cut of $i-z$ $>$ 0.7\,mag as a criterion to identify new UCDs. It includes almost all stars of subtype M7\,V and later, at the expense of some contamination from earlier subtypes (M2--6\,V). To minimize this contamination as much as possible, we used VOSA\footnote{\tt http://svo2.cab.inta-csic.es/theory/vosa/} \citep[VO Sed Analyser;][]{Bayo08} to derive the effective temperature of the photometrically selected candidate UCDs (i.e. objects with $i-z$ $>$ 0.7\,mag). VOSA is a tool developed by the Spanish Virtual Observatory designed to estimate physical parameters from the comparison of the observed target SED to different collections of theoretical models (see an illustrative example in Fig.\,\ref{fig:vosa}).

\begin{table}
\centering
	\caption{Number of M dwarfs in \citet{West11} with $i$, $z$ photometry in J-PLUS.}
	\label{tab:teffcol}
	\begin{tabular}{cc} % four columns, alignment for each
		\hline
		\hline
		\noalign{\smallskip}
		Spectral type & Number of objects \\
		\hline
			\noalign{\smallskip}
		M0\,V & 111 \\ 
		M1\,V & 79 \\ 
		M2\,V & 131 \\
        M3\,V & 136 \\
        M4\,V & 119 \\
        M5\,V & 92 \\
        M6\,V & 298 \\
        M7\,V & 239 \\
        M8\,V & 39 \\
        M9\,V & 13 \\
        \noalign{\smallskip}
		\hline
	\end{tabular}
\end{table}

Observational SEDs covering the optical and infrared range were built using the photometric catalogues available in VOSA. Then, SEDs were compared to the grid of BT-Settl CIFITS model atmospheres \citep{Baraffe15} to estimate effective temperatures. We assumed $\log{g}$\,$\geq$\,4.5 and solar metallicity. {\it GALEX} photometry was not included in the SED as, for  M dwarfs,  most of the  flux  in  the {\it GALEX} bandpasses does not have a photospheric origin but chromospheric \citep{Shk11}. Consequently, the flux observed at {\it GALEX} bands is systematically higher than the flux estimated from the photospheric models, which would have a clear impact on the SED fitting. 

Extinction can play an important role in shaping the SED and thus, in the estimation of physical parameters. To account for this effect, we decided to leave extinction as a free parameter in the SED fitting process with values ranging from $A_V$= 0.0\,mag to 0.7\,mag.
%Although our targets are located at relatively  high  Galactic latitudes (Figure\,\ref{fig:aladin}) where the effect of reddening due to interstellar dust should not be large, 

Using the 70\,840 spectroscopically confirmed M dwarfs available in \cite{West11}, we derived a relationship between spectral types and effective temperatures calculated using VOSA and \cite{Baraffe15} models (Table\,\ref{tab:teffspt}). According to this, we adopted for UCDs a conservative value of \teff\,$\leq$\,2900\,K. With this criterion we may loose a small fraction of hotter UCDs but, at the same time, we avoid contamination from a significant number of M5--6 dwarfs.

\begin{table}
	\centering
	\caption{Mean effective temperatures calculated using VOSA and BTSettl-CIFITS models as a function of the spectral type given by \citet{West11}.}
	\label{tab:teffspt}
	\begin{tabular}{ccc} % four columns, alignment for each
		\hline
		\hline
		\noalign{\smallskip}
		Spectral & Number of & \teff \\
        type     & objects  &  [K]  \\
		\hline
		\noalign{\smallskip}
		M5\,V & 3\,901 & 3\,210$\pm$210\\
		M6\,V & 5\,645 & 3\,060$\pm$110\\
		M7\,V & 5\,824 & 2\,960$\pm$150\\
        M8\,V & 1\,682 & 2\,710$\pm$160\\
        M9\,V & 891  & 2\,600$\pm$160 \\
        \noalign{\smallskip}
		\hline
	\end{tabular}
\end{table}

SEDs of the photometrically selected ultracool candidates with \teff\,$\leq$\,2900\,K were visually inspected to remove bad photometric points (i.e. points that clearly deviate from the theoretical SED) that could be affecting the quality of the SED fitting and, thus, the estimation of the effective temperature. Moreover, we used Aladin \citep{Bonnarel00} to inspect the associated images and discard any problem related to blending or contamination by nearby sources. 

The next step was to place our photometrically selected candidate UCDs in a colour-magnitude diagram. To do so, we reproduced the diagram shown in Fig. 6 of \cite{Babu18} with all the {\it Gaia} DR2 objects at less than 100 pc. Among our candidates, we kept only those with relative errors of less than 10\% in $G$ and G$_{RP}$ and less than 20\% in parallax. 
The absolute {\it Gaia} magnitude in the $G$ band for individual stars was estimated using
\begin{equation}
  M_G = G + 5 \log{\varpi} +5,   
\end{equation}
where $\varpi$ is the parallax in arcseconds. In our case, the inverse of the parallax is a reliable distance estimator as we are keeping only sources with relative errors in parallax lower than 20\% \citep{Luri18}. 

\begin{figure}
	\includegraphics[width=\columnwidth]{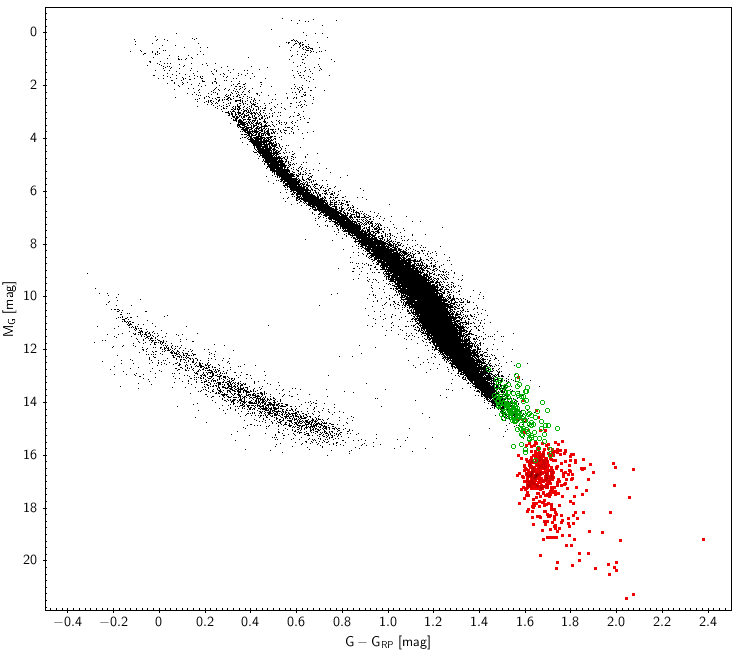}
    \caption{Location of our photometrically selected candidate UCDs (open green circles) in a colour-magnitude diagram built using {\it Gaia} DR2 sources with parallaxes larger than 10 milliarcseconds (black dots). L and T dwarfs identified in \citet{Smart17} and with counterparts in {\it Gaia} DR2 are overplotted in red.
}
	\label{fig:cand_phot}
\end{figure}

We also forced that all {\it Gaia} counterparts have a value of {\tt visibility$_{-}$periods$_{-}$used} > 6, which is the minimum value to accept a five-parameter astrometric solution for {\it Gaia} sources \citep{Lindegren18}. This keyword indicates the number of groups of observations separated from other
groups by a gap of at least four days \citep{Babu18}. After all this filtering process, we ended up with 127 new potential candidate UCDs not previously reported in the literature. Figure\,\ref{fig:cand_phot} shows the position of these sources in the CMD diagram. It confirms the dwarf and cool nature of all our candidates. 

\subsection{Kinematic search}
\label{kinematic}
Proper motions have been widely used in the literature to distinguish M dwarfs from other objects with similar colours but different kinematics, such as M giants, quasars, or distant red luminous galaxies \citep[e.g.,][]{Theissen17, Theissen16, Caballero08}. Moreover, proper motions are very useful to identify ultracool dwarfs with atypical colours due to anomalous values of gravity \citep{Schmidt10} or metallicity \citep{Bochanski13, Zhang18} that would be missed by colour cuts used in photometry-only searches. 

The J-PLUS limiting magnitudes in the $i$- and $z$-band, together with the UCD colour-distance relationship provided in \cite{Theissen17}, gives a maximum distance of about 400\,pc to find UCDs in the J-PLUS survey. This implies that UCDs seen by J-PLUS are at distances where proper motions are significant. 

\begin{figure}
\includegraphics[width=\columnwidth]{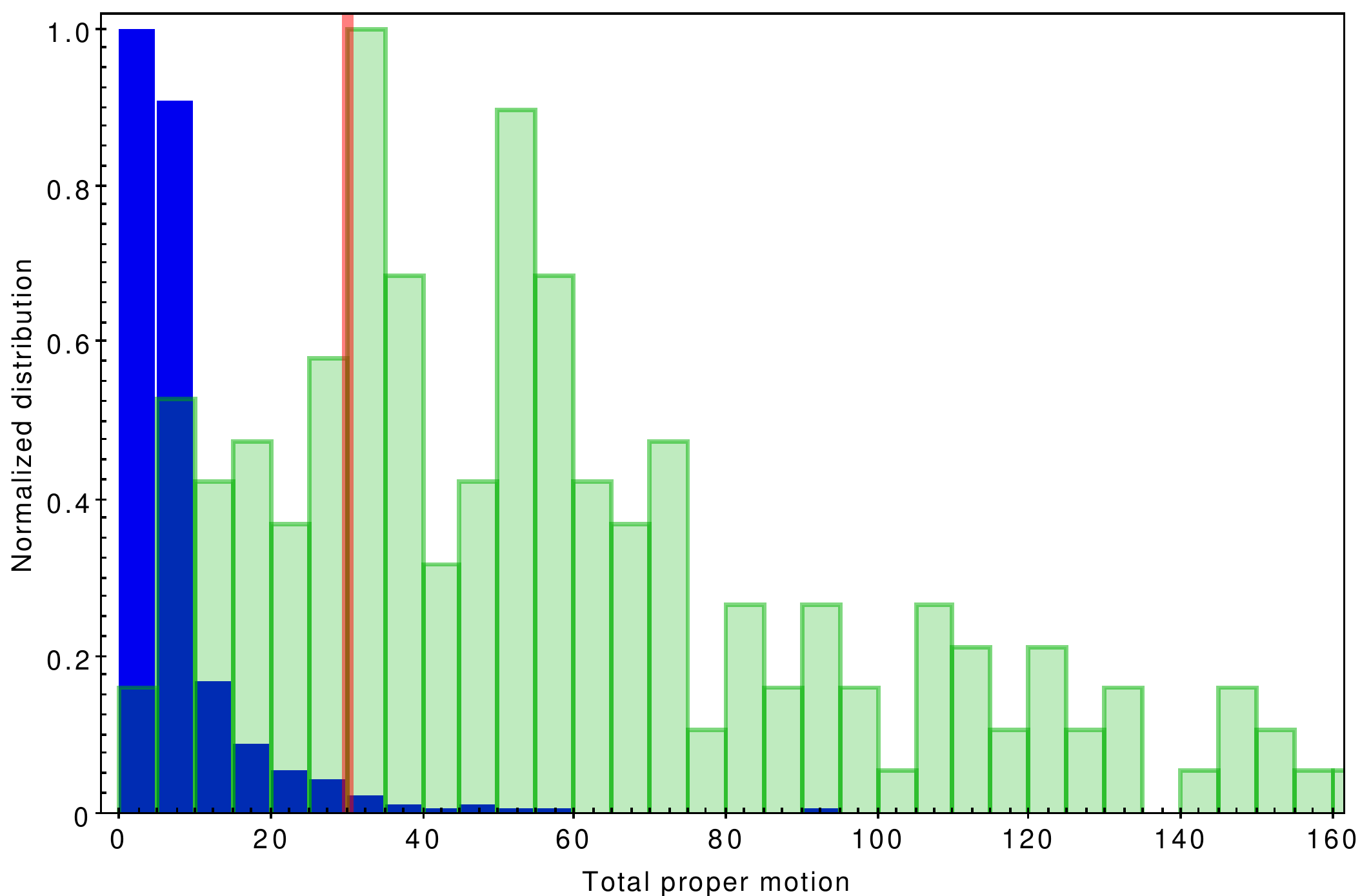}
\caption{Normalized distribution functions of the {\it Gaia} DR2 proper motions of the subgiants/giants available in SIMBAD with spectral type later than M5 (blue) and the 127 candidate UCDs identified in Sect.\,\ref{phot} (green). The cut at 30\,mas/yr is marked with a vertical red line.}
   \label{fig:pm_histo}
\end{figure}

To define the proper motion cut between M dwarfs and M subgiants/giants, we took all objects classified in SIMBAD\footnote{\tt http://simbad.u-strasbg.fr/simbad/} \citep{Wenger00} as M subgiant/giant (luminosity classes IV, IV/III, and III) and with spectral type later than M5 (1\,093 objects), and looked for their proper motions in {\it Gaia} DR2. Their normalized total proper motion distribution is compared in Fig.\,\ref{fig:pm_histo} with that of the 127 candidate UCDs identified in Sect.\,\ref{phot}. A proper motion cut at 30\,mas/yr represents a good balance between giant contamination and dwarf completeness.

In the kinematic search we limited our query in J-PLUS to {\tt class\_star} $\geq$ 0.5. Compared to the photometric search we have relaxed this condition from 0.8 to 0.5 based on two reasons. First, it increases the number of objects to be studied. Second, the contamination of non-stellar objects wrongly flagged with {\tt class\_star} $>$ 0.5 can be easily removed using the proper motion information as extragalactic objects have negligible spatial motions. Sources fulfilling this condition were crossmatched with the {\it Gaia} DR2 catalogue to get the information on proper motions. A search radius of 10\,arcsec was used. If more than one counterpart exist in that radius we took the nearest one. Of the 2\,620\,881 J-PLUS sources with {\it Gaia} DR2
counterparts, 64\,134 have total proper motions larger than 30\,mas/yr. 

\begin{figure}
\includegraphics[width=\columnwidth]{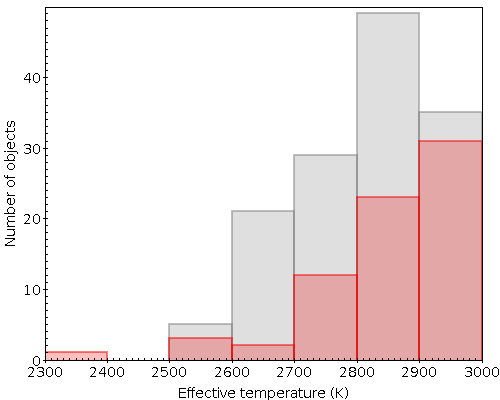}
\caption{\teff$ $ distribution of the candidate UCDs identified using photometry (grey) and proper motions (red). Most of our candidate UCDs ($\sim$ 65 \%) have \teff $>$ 2800\,K. }
   \label{fig:teff}
\end{figure}

To remove hotter objects (spectral types earlier than M7) with high proper motions, we estimated effective temperatures for the 64\,134 objects using VOSA following a procedure similar to that described in Sect.\,\ref{phot}. Also, objects with {\it Gaia} counterparts having {\tt visibility$_{-}$periods$_{-}$used} < 6 and  {\tt astrometric$_{-}$excess$_{-}$noise} > 1 were removed. The latter is a {\it Gaia} parameter that quantifies the scatter of residuals in the astrometric solution. According to \cite{Lindegren18}, values above 1 will, most likely, correspond to artefacts. After this filtering, 60 new UCDs candidate were kept. These objects were not identified in the photometric search mainly because of two reasons: J-PLUS {\tt class$_{-}$star} $\leq$ 0.8 or bad photometry in the J-PLUS $i$-, $z$-bands.The distribution of effective temperatures of our candidate UCDs (identified both using photometry and proper motions) are given in Fig.\,\ref{fig:teff}.

\begin{figure}
\includegraphics[width=\columnwidth]{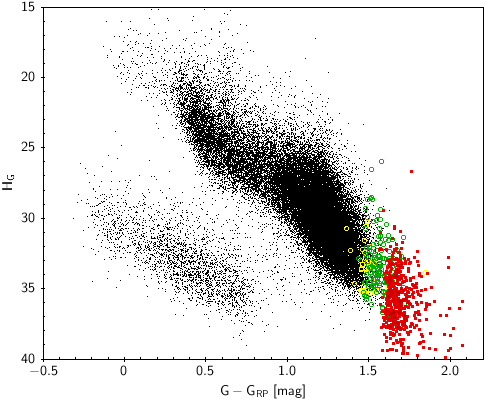}
\caption{Reduced proper motion - colour diagram. In black we plot the objects used in Fig.\,\ref{fig:cand_phot}. Green open circles represent the 127 candidate UCDs identified in Sect. \ref{phot}. Yellow open circles indicate the 60 kinematically selected candidate UCDs. L and T dwarfs from \citet{Smart17} used in Fig.\,\ref{fig:cand_phot} are plotted as red squares.}
   \label{fig:rpm}
\end{figure}

Figure\,\ref{fig:rpm} shows the position of the 60 kinematically-selected candidate UCDs in a reduced proper motion diagram:
\begin{equation}
H_{G} = G + 5 \log{\mu} + 5
\end{equation}
where $G$ is the {\it Gaia} magnitude and $\mu$ is the total proper motion in mas yr$^{-1}$. In this type of diagrams, proper motion is used as a proxy for distance assuming that nearby objects will have larger proper motions. Just as in a Hertzsprung-Russell diagram, reduced proper motion diagrams have demonstrated to be excellent tools to segregate the various stellar populations 
\citep{Dhital10, Jimenez11, Jimenez12, Zhang13, Smart17}. All our candidates lie in the expected locus for ultracool dwarfs.

The approach adopted in this work for the kinematic search has two main limitations. First, we keep only objects having a minimum total proper motion $\mu$ $>$ 30 mas/yr, which may exclude a non-negligible amount of nearby objects. This can be seen in Fig.\,\ref{fig:pm_histo}, where we favour low degree of contamination to completeness.  Second, our procedure is based on {\it Gaia} proper motions. {\it Gaia} operates at optical wavelengths with a limiting magnitude $G$ $\sim$ 21.4\,mag, which makes it incomplete for UCDs, in particular for the reddest objects, as they emit predominantly in the infrared and are very faint in the {\it Gaia} bands \citep[e.g.,][]{Smart17}. An alternative to this approach would be the estimation of proper motions from the positions in optical and infrared surveys taken at different epochs, a methodology that is beyond the scope of this exploratory paper. Moreover, using tangential velocities, where information on distances and proper motions is combined, could be an alternative approach to separate dwarfs from giants. However, the number of {\it Gaia} DR2 parallaxes accurate enough to provide unbiased estimations of distances is significantly lower than the number of accurate proper motions (i.e., relative errors below 10\% in both components). Using tangential velocities would result in the loss of 20-25\% of our kinematically selected candidates which led us not to use them in our kinematic search of UCDs. 

A sample of the photometrically (127) and kinematically (60) selected candidate UCDs is given in Table\,\ref{tab:cand}, while the full list of candidates is available online at the SVO-JPLUS archive of ultracool dwarfs (see Appendix A). This service will be updated and maintained in the framework of the Spanish Virtual Observatory\footnote{\tt http://svo.cab.inta-csic.es}. According to VOSA criteria, based on the slope of the SED and the flux difference between the observed and theoretical SEDs (see the VOSA help\footnote{\tt https://bit.ly/2yzwyIF} for a detailed description on how VOSA handles the infrared excess), none of the candidate UCDs show excess in the infrared that could be ascribed to circumstellar material or to the presence of a close ultracool companion. This result is in agreement with the ratio of M dwarfs with infrared excess given by \cite{Theissen14}, estimated in one object every about 400 stars.

\begin{table*}
	\centering
	\caption{New candidate ultracool dwarfs.}
	\label{tab:cand}
	\begin{tabular}{cccccrcccc} % four columns, alignment for each
		\hline
		\hline
		\noalign{\smallskip}
		$\alpha$ & $\delta$ &  \teff & G& $\varpi$ & $\mu_{\alpha}\cos{\delta}$ & $\mu_{\delta}$ & Search & $i$ & $z$ \\
       (ICRS, deg) & (ICRS, deg) & [K] &[mag] & [mas] & [mas/yr] & [mas/yr] & &[mag] &[mag] \\
       \hline
       \noalign{\smallskip}
       21.70712& 33.44938 & 2600 & 20.94 & 11.18 &  7.36& -51.33  & P & 20.19 & 18.69\\
  22.16372 & 33.30222 & 2600 & 18.44 & 18.93 & 77.34  & -99.44 &  P & 17.81 & 16.33\\
  23.15882& 39.59608 & 2800 & 20.58 & 3.41 & -54.42  & -21.40 & K &  & \\
  27.63937 & 33.73132& 2900 & 19.84 & 3.79 & 2.44 & -39.94 & P & 19.34& 18.11\\
  28.13097 & 33.83401 & 2600 & 19.69 & 11.93  & 58.33   & -46.75  & P & 19.11 & 17.67\\
  ... & & & & & & & & & \\
  \noalign{\smallskip}
		\hline

	\end{tabular}
	\tablefoot{Equatorial coordinates, $G$ magnitudes, parallaxes and proper motions are from {\it Gaia} DR2, effective temperatures are estimated with  VOSA and $i$,$z$ magnitudes are from J-PLUS. The column ``Search'' indicates whether the source was discovered in the photometric (P) or in the kinematic (K) search. The full table is available at the SVO-JPLUS archive of ultracool dwarfs (see Appendix A).}
\end{table*}

\subsection{Known ultracool dwarfs}
In this section we assess the fraction of known UCDs that have been recovered using our methodology. In particular we looked for UCDs in two catalogues: SIMBAD and \cite{Smart17}. 

Using the SIMBAD TAP service\footnote{\tt http://simbad.u-strasbg.fr:80/simbad/sim-tap}, we chose among the more than 9 million available objects those having M7V, M8V, M9V, and L spectral types. A total of 9\,894 objects fulfilled these conditions. 

To know how many of them lie in the region of the sky covered by IDR201709, we took advantage of the Multi-Object Coverage\footnote{\tt http://ivoa.net/documents/MOC/index.html} VO standard and Aladin. A total of 588 objects lie in the region of the sky observed in IDR201709.  A query to the archive for these 588 objects returned 368 objects with J-PLUS data. We repeated the same methodology for \cite{Smart17}. Only 53 out of the 1885 objects included in the catalogue lie in the J-PLUS field, 7 of them with observations in the J-PLUS archive. The scarcity of J-PLUS counterparts can be explained by the intrinsic faintness of these objects, typically beyond the IDR201709 limiting magnitude. 

The efficiency of our search was estimated using the false negative rate (number of known UCDs that were not rediscovered in our search). All SIMBAD sources (368) were found while only four of the seven objects included in \cite{Smart17} with J-PLUS observations were recovered, the remaining three having a bad SED fitting that prevented us from deriving their effective temperatures. These 372 (368 + 4) objects were not included in our list of 187 (127 + 60) objects as they have already been reported in the literature. This gives a total of 559 UCDs identified in this paper. The 187 new candidate UCDs compared to the 9\,894 UCDs already included in SIMBAD represents a 2\% of new discoveries. However, if we restrict our analysis to the area of the sky covered by J-PLUS DR1 (528\,deg$^2$), the increase in the number of known UCDs is about 50\% (187/372). This ratio of new discoveries is not evenly distributed with spectral types, being higher ($\sim$ 5\%) for the cooler (> M8) objects.  Extrapolating these results to the foreseen sky coverage at the end of the J-PLUS survey (8\,500 deg$^{2}$) we would expect to discover $\sim$ 3\,000--3\,500 new UCDs, which represents one third of the UCDs presently included in SIMBAD. 

These results confirm the robustness of our methodology to discover ultracool dwarfs.

\section{Results and discussion}

\subsection{Distances}

\begin{figure}
\includegraphics[width=\columnwidth]{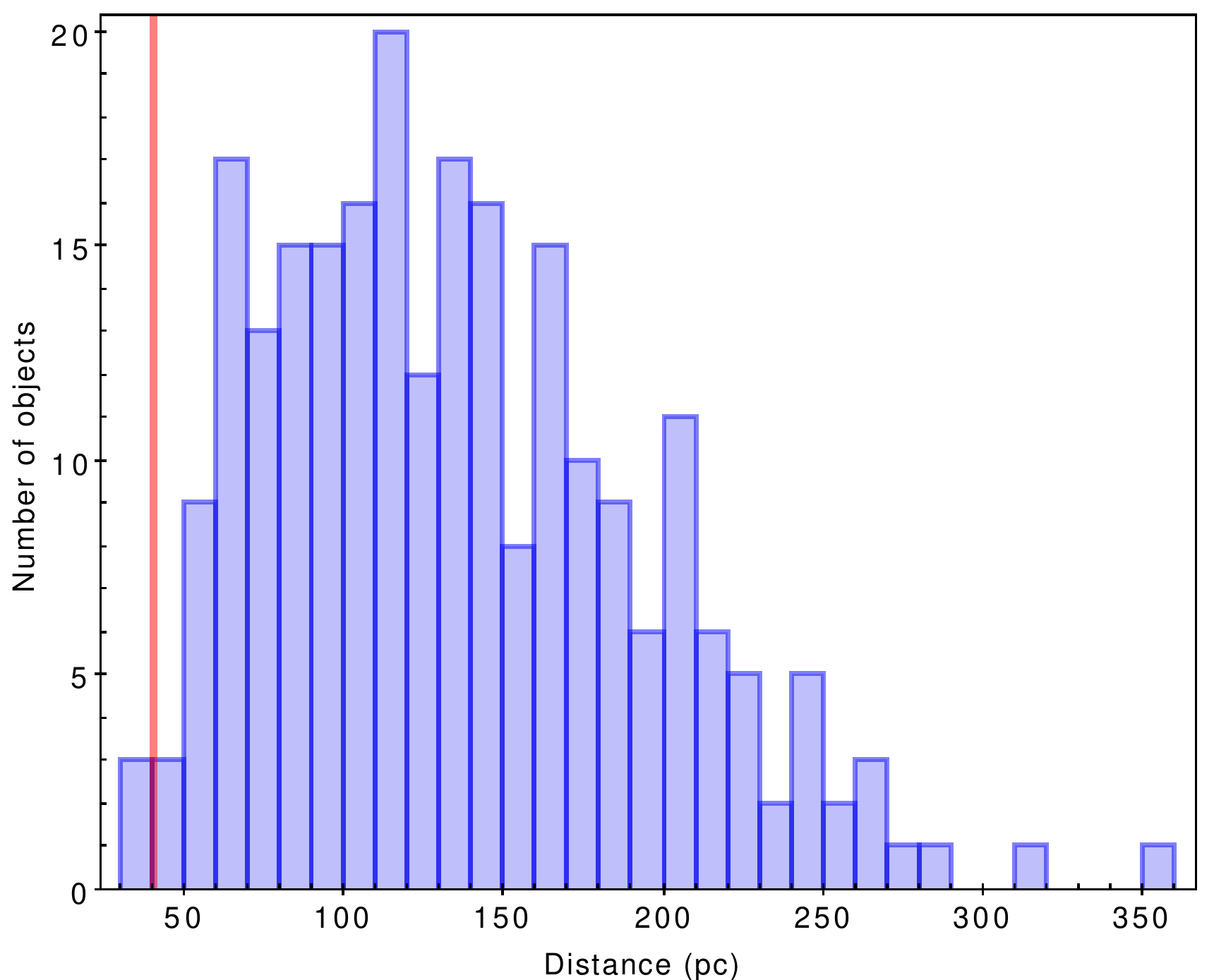}
\caption{Distance distribution for our candidate UCDs with good {\it Gaia} DR2 parallaxes ($\delta \varpi / \varpi$ $<$ 20\%). The red vertical line indicates d = 40\,pc. The mean value of the distribution is 130 pc, with the closest and
farthest objects at 32 pc and 352 pc.}
\label{fig:dist}
\end{figure}

To avoid biases due to large errors in parallax \citep{Astra2016}, distances were estimated only if the relative error in parallax was less than 20\%. As in Sect.\,\ref{phot}, we used the inverse of the parallaxes given in {\it Gaia} DR2. The distribution of
distances for our candidate UCDs is shown in Fig.\,\ref{fig:dist}.

%Nearby ultracool dwarfs are ideal candidates for numerous astrophysical investigations like, for instance, the study of moving groups, extrasolar planets or close and wide binaries \citep{Theissen17}.

Due to the comprehensive search done in the past for nearby UCDs, the great majority of the new candidate UCDs presented in this paper are at distances greater than 40 pc, which is interesting as reference for future wide and deep surveys such as Euclid \citep{Racca18}. However, there are three objects at less than 40 pc that escaped from previous searches, which indicates that there is still room to discover new nearby UCDs even with small telescopes such as JAST/T80. We list the details of these three objects in Table\,\ref{tab:nearby}. Based on the effective temperatures calculated with VOSA and the \teff -- spectral type relationship given in (Table\,\ref{tab:teffspt}), we estimated a photometric spectral types of M7 for \teff\, $\sim$ 2900\,K and M8 for \teff\, $\sim$ 2700\,K.

\begin{figure}
\includegraphics[width=\columnwidth]{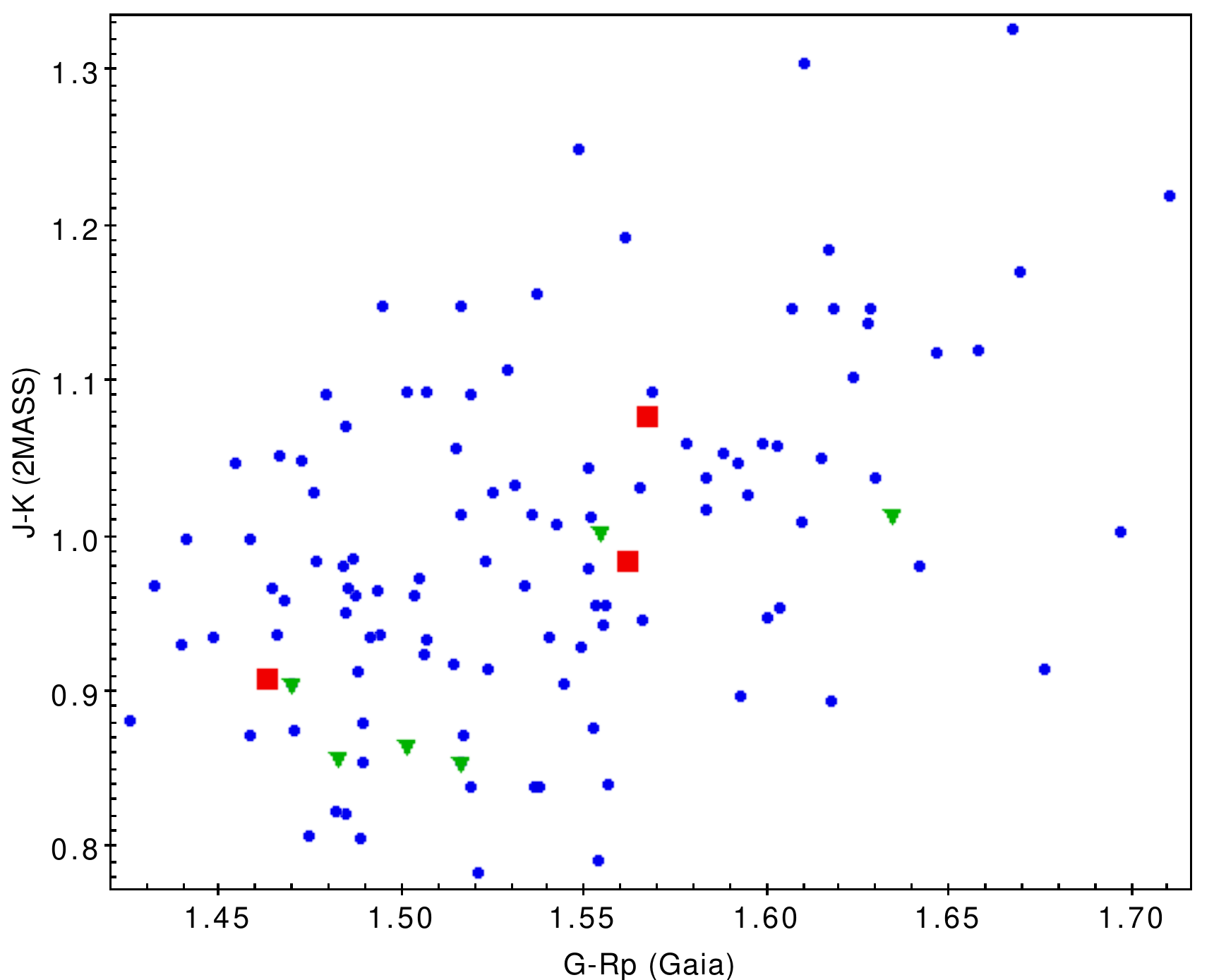}
\caption{$J-K_{s}$ vs $G-Rp$ colour-colour diagram of our candidate UCDs with good 2MASS photometry (Qflg=A) in $J$ and $K_{s}$ bands. Red squares represent our three candidate UCDs at distances $d$ $<$ 40 pc. Inverted green triangles indicate objects with tangential velocities V$_{\rm tan}$  $>$ 100 km s$^{-1}$ }
\label{fig:JK}
\end{figure}

Ultracool dwarfs may escape from photometric surveys because their unusual $J-K_{s}$ colours for their spectral types. Red $J-K_{s}$ outliers often exhibit distinctive low-gravity features, indicating that these objects are young and still collapsing. On the other hand, blue outliers show strong H$_{2}$O and weak CO features in their infrared spectra, possibly associated to high gravity and old age \citep{Burgasser08, Schmidt10b, Zhang17}. A paradigmatic example of these outliers is SDSS J141624.09+134826.7, an exceptionally blue L dwarf within 10 pc of the Sun \citep{Bowler10,Schmidt10}. Nevertheless, as shown in Fig.\,\ref{fig:JK}, this was not the case of our three candidate UCDs, which lie in the region of typical colours. 

\begin{table*}
	\centering
	\caption{Newly identified nearby ($d$ $<$ 40 pc) candidate UCDs. Coordinates are from {\it Gaia} DR2. We adopted the step of the BT-Settl CIFITS grid of models (100\,K) as the error in effective temperatures.}
	\label{tab:nearby}
	\begin{tabular}{ccc ccccc} % four columns, alignment for each
		\hline
		\hline
		\noalign{\smallskip}
		{\it Gaia} ID& $\alpha$ & $\delta$ & $G$ & \teff & $\mu_{\alpha}\cos{\delta}$ & $\mu_{\delta}$ & $d$ \\
       DR2 & (deg) & (deg) & [mag] & [K] & [mas/yr] & [mas/yr] & [pc] \\
		\hline
		\noalign{\smallskip}
1902388491693623680 & 338.44505& 33.99409& 17.57 & 2900& 254.44$\pm$ 0.25& 104.42$\pm$0.26 & 32.45$\pm$0.15 \\
2109889524381399040 & 274.94447&41.03144& 17.89 &  2700&21.30$\pm$0.23&-250.82$\pm$0.24& 39.60$\pm$0.19 \\
1596590634148306944 & 236.27034& 52.23673 & 17.16 & 2900 & 57.26$\pm$0.18 & 86.54$\pm$0.18& 39.98$\pm$0.15 \\
\noalign{\smallskip}
		\hline
	\end{tabular}
\end{table*}

\subsection{Tangential velocities}

Kinematics is a good proxy to get information on the ages of UCDs as old stars belong to the thick disk and, thus, have larger velocities than young stars, which belong to the thin disk \citep{Montes01, Zuckerman04}.

%As a stellar or substellar object becomes older, the number of gravitational interactions increases, changing its velocity in random directions and moduli. Thus, the velocity dispersion for older population is larger than for younger populations \citep{Schmidt10b}.

The lack of radial velocities for our ultracool candidates prevents from making a three dimensional kinematic analysis. Instead, by combining {\it Gaia} DR2 proper motions and distances, we used tangential velocities, defined as V$_{\rm tan}$ = 4.74 $\mu$ $d$, where V$_{\rm tan}$ is given in km s$^{-1}$, $\mu$ is the total proper motion in arcsec yr$^{-1}$ and $d$ is the distance in pc.  

\begin{figure}
\includegraphics[width=\columnwidth]{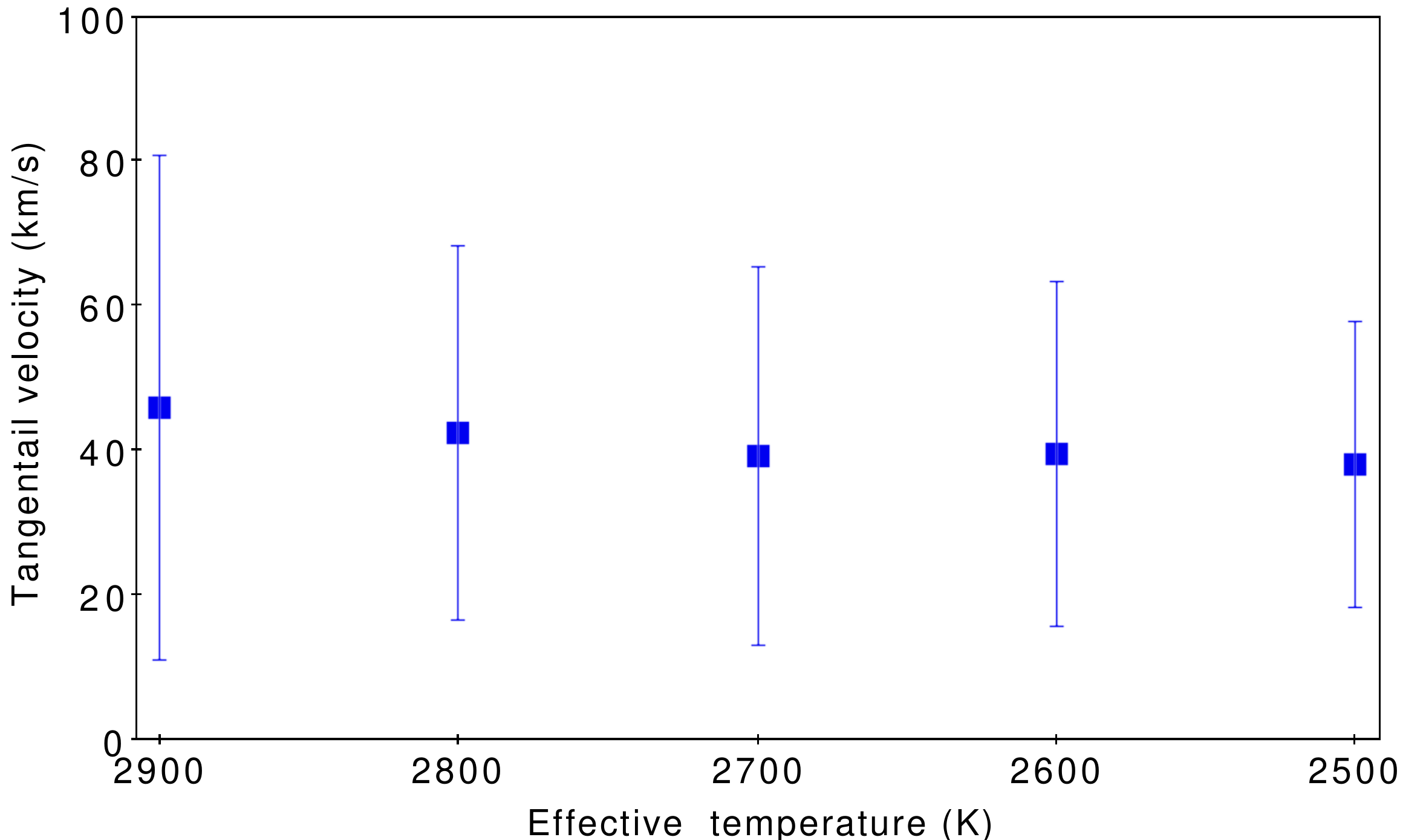}
\caption{Mean tangential velocities of our candidate UCDs. Error bars represent the standard deviation.}
\label{fig:vtspt}
\end{figure}

Figure\,\ref{fig:vtspt} shows the distribution of the mean tangential velocities with the effective temperature. The discretization seen in the values of \teff$ $ is a consequence of the step adopted in the CIFITS grid of models. The lack of correlation between tangential velocities and effective temperature indicates that our candidates represent a single kinematic population irrespective of the spectral type. 

\begin{figure}
\includegraphics[width=\columnwidth]{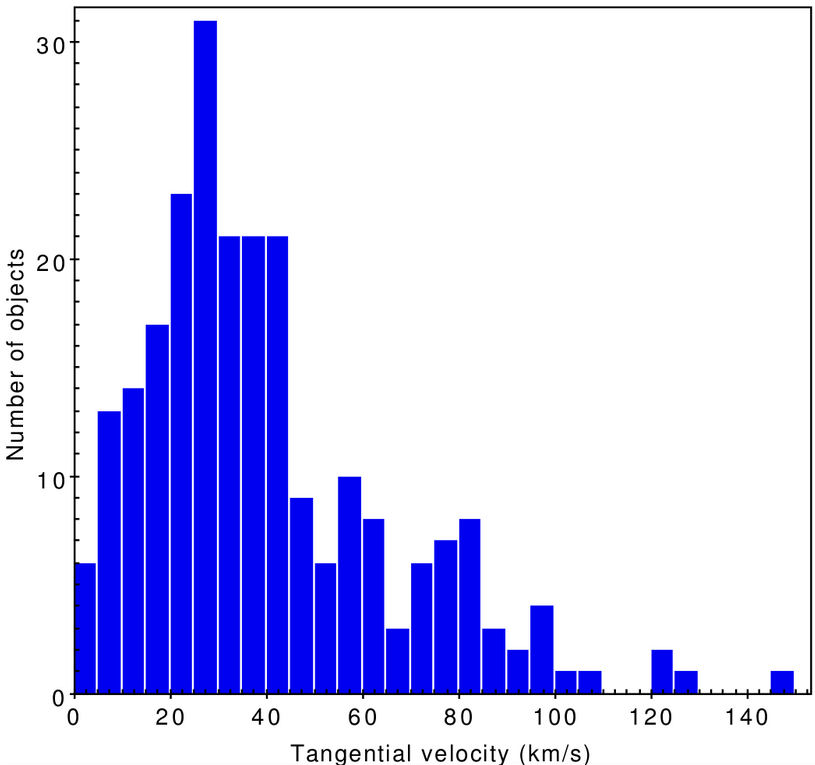}
\caption{Histogram of tangential velocities of our candidate UCDs}
\label{fig:vtan}
\end{figure}

Figure\,\ref{fig:vtan} shows the distribution of tangential velocities. There
are six objects with \vt > 100 km s$^{-1}$. Even considering these objects we obtain a median value of \vt= 34 km s$^{-1}$ and a dispersion of $\sigma$ = 26 km s$^{-1}$, which coincides very well with previous determinations for ultracool dwarfs \citep{Faherty09}. 

These six high-velocity objects do not show unusual colours for their spectral type (Fig.\,\ref{fig:JK}). Adopting \vt = 180 km s$^{-1}$ as the boundary between disk and halo objects \citep{Sesar2008}, we conclude that all our UCD candidates correspond to thin disk objects, with the six high-velocity objects being potential members of the thick disk.

\section{Conclusions}

Using a Virtual Observatory methodology we have produced a catalogue of 187 new candidate ultracool dwarfs using J-PLUS IDR201709. Our goal was not to construct a complete catalogue. Rather, we attempted to build a catalogue of bona fide UCDs and to test a search methodology that could be used for new J-PLUS releases and, what is more interesting, for the deeper J-PAS survey with first light happening in 2019. J-PAS, with its contiguous system of 54 narrow-band filters, will provide an unprecedented photometric coverage in the optical range, allowing more accurate determinations of the effective temperatures and, probably, a direct confirmation of the true UCD nature of our objects without the need for external spectroscopic follow-up.   

The use of a dual methodology based on photometry and proper motions tends to minimize the drawbacks and biases associated to the search of ultracool objects: photometric-only selected samples may leave out peculiar UCDs not following the canonical trend in colour-colour diagrams, while proper motion searches may ignore objects with small values of projected velocity in the plane of the sky. The high success ratio recovering known UCDs ($>95\%$) demonstrates the robustness of our procedure.

With the help of the VOSA Virtual Observatory tool we estimated effective temperatures for our candidate UCDs. They range from M7\,V to L0\,V with the great majority of the objects having M7--8\,V spectral types. Distances were computed using {\it Gaia} parallaxes with relative errors less than 20\%. A mean value of 130 pc was obtained, with the closest and
farthest objects at 32 pc and 352 pc, respectively. We also identified three new systems with distances $d <$ 40 pc.

Analysis on tangential velocities concluded that our candidate UCDs are consistent with the Galactic thin-disk population. However, there are six objects that lie at
the tail end of the velocity distribution and are likely part of an older Galactic population. 

%With the new J-PLUS and the deeper J-PAS releases, these surveys will become very valuable resources to identify and characterize ultracool dwarfs.

\begin{acknowledgements}

We thank our referee, Dr. Zenghua Zhang, for his constructive suggestions. We deeply thank J-PLUS collaboration, in particular, S. Akras, A. Alvarez, P. Coelho, and V. Placco for constructive comments that helped to improve the presentation of
the results. We acknowledge use of the ADS bibliographic services. This research has made use of the Aladin sky atlas developed at CDS, Strasbourg Observatory, France. This publication makes use of VOSA, developed under the Spanish Virtual Observatory project supported from the Spanish MINECO through grant AyA2017-84089. This  work  has  made  use  of  data  from  the  European
Space Agency (ESA) mission
{\it Gaia}\footnote{\tt https://www.cosmos.esa.int/gaia}, processed by the
{\it Gaia} Data Processing and
Analysis  Consortium  (DPAC)\footnote{\tt https://www.cosmos.esa.int/web/gaia/dpac/consortium}. We also extensively made use of the Vizier and SIMBAD services, both operated at CDS, Strasbourg, France.

Funding for the J-PLUS Project has been provided by the Governments of Spain and Arag\'{o}n through the Fondo de Inversiones de Teruel, the Spanish Ministry of Economy and Competitiveness (MINECO; under grants AYA2015-66211-C2-1-P, AYA2015-66211-C2-2, AyA2012-30789 and ICTS-2009-14), and European FEDER funding (FCDD10-4E-867, FCDD13-4E-2685). The Brazilian agencies FAPERJ and FAPESP as well as the National Observatory of Brazil have also contributed to this project. RAD acknowledges support from CNPq through BP grant 312307/2015-2, CSIC through grant COOPB20263, FINEP grants REF. R
1217/13 - 01.13.0279.00 and REF 0859/10 - 01.10.0663.00 for hardware support for the J-PLUS project through the National Observatory of Brazil. E. L. Mart\'{\i}n acknowledges support from project AYA2015-69350-C3-1-P. J. A. Hernandez-Jimenez thanks to Brazilian  institution CNPq  (project 150237/2017-0) and
Chilean institution CONICYT through Programa de Astronom\'{\i}a, Fondo ALMA-CONICYT 2017 (project 31170038). R.~Lopes was partially supported by the Brazilian agency CNPq (PQ 302037/2015-2 and PDE 200289/2017-9). F. J. E. acknowledges financial support from the ASTERICS project (ID:653477, H2020-EU.1.4.1.1. Developing new world-class research infrastructures). This   work   is   based   on   observations   made   with   the
JAST/ T80  telescope  at  the  Observatorio  Astrofísico  de  Javalambre (OAJ) in Teruel, owned, managed and operated by the Centro de Estudios de Física del Cosmos  de  Aragón. We thank the OAJ Data Processing and Archiving Unit (UPAD) for
reducing and calibrating the OAJ data used in this work.

\end{acknowledgements}

%-------------------------------------------------------------------

\bibliographystyle{aa}
\bibliography{ref_ucd.bib}

\appendix

\section{Virtual Observatory compliant, online catalogue}

In order to help the astronomical community on using our
catalogue of candidate UCDs, we developed an archive system  that  can  be  accessed  from  a  webpage\footnote{\tt http://svo2.cab.inta-csic.es/vocats/v3/ucd\_jplus/index.php} or  through  a
Virtual Observatory ConeSearch\footnote{\tt (e.g. https://bit.ly/2PuqlJ5)}.

The  archive  system  implements  a  very  simple  search
interface that allows queries by coordinates and radius as
well as by other parameters of interest. The user can also select the maximum number of sources (with values from 10 to
unlimited) and the number of columns to return (minimum,
default, or maximum verbosity).
The result of the query is a HTML table with all the
sources found in the archive fulfilling the search criteria. The
result can also be downloaded as a VOTable or a CSV  file.
Detailed information on the output  fields can be obtained
placing the mouse over the question mark located close
to the name of the column. The archive also implements the
SAMP\footnote{\tt http://www.ivoa.net/documents/SAMP}
(Simple  Application  Messaging)  Virtual  Observatory  protocol.  SAMP  allows  Virtual  Observatory  applications  to  communicate  with  each  other  in  a  seamless  and
transparent manner for the user. This way, the results of a
query  can  be  easily  transferred  to  other  VO  applications,
such as, for instance, Topcat.

\end{document}